\documentclass[reprint,aps,eqsecnum,twoside]{revtex4-2}

\usepackage{hyperref,amsfonts,amsmath,amssymb,graphicx,bm}

\begin{document}

\author{Maxim Dvornikov}
\email{maxim.dvornikov@gmail.com}

\title{Quantization of massive Dirac neutrinos in external fields}

\affiliation{Pushkov Institute of Terrestrial Magnetism, Ionosphere
and Radiowave Propagation (IZMIRAN),
108840 Moscow, Troitsk, Russia}

\begin{abstract}
We review the applications of the quantum field theory (QFT) for the
description of massive Dirac neutrinos in external fields. Two particular
cases of external background are considered. First, we examine neutrinos
in background matter. Then, we study neutrinos with anomalous magnetic
moments in a magnetic field. In both situations, we derive the operator
valued neutrino wavefunctions, accounting for external fields, which
obey the canonical anticommutation relations. Then, we check that
the total energy and momentum of a neutrino field have the appropriate
forms. Using the exact solution in a Dirac equation in a magnetic field, we also derive the propagator for a massive Dirac neutrino in this external background. The results obtained are of importance for the QFT application
to neutrino oscillations in external fields.
\end{abstract}

\maketitle

\section{Introduction}

The neutrino interaction with external fields can significantly modify
the process of neutrino flavor oscillations. The probability of neutrino
oscillations was predicted in Ref.~\cite{MikSmi85} to be significantly
amplified if neutrinos interact electroweakly with background matter.
This fact underlies the Mikheyev--Smirnov--Wolfenstein effect~\cite{MikSmi85,Wol78}
which is a plausible explanation of the solar neutrino problem~\cite{FogLis04}.

Another external field which can affect neutrino oscillations is the
neutrino electromagnetic interaction. Despite the electric charge
of a neutrino is likely to be zero, these particles can have magnetic
moments~\cite{LeeSch77}. If such a neutrino interacts with the magnetic
field, its spin can precess~\cite{Cis71}. Hence the conversion of
a left-handed active neutrino to a right-handed sterile particle is
possible. Moreover, neutrinos can have transition magnetic moments
which mix different neutrino types. In this situation, the neutrino
spin-flavor precession can happen~\cite{LimMar88,Akh88}.

Thus, the study of neutrino interactions with external fields is of
great importance for the description of neutrino oscillations. Moreover,
since our world is generically based on quantum principles, the properly
quantized neutrino fields interacting with external backgrounds are
desirable. For example, these quantized neutrino fields can be utilized
in the quantum field theory based approach for the description of
neutrino oscillations~\cite{Beu03,NauNau20}.

The quantum field theory of a massive neutrino in an external field
involves the exact solution of the wave equation in the field in question. The
solution of the Dirac equation for a neutrino in matter was found in Refs.~\cite{StuTer05,Lob05}.
The neutrino quantum states in rotating matter were described in Ref.~\cite{GriSavStu07}.
The effect of the matter rotation on neutrinos by solving the Dirac
equation in a noninertial frame was discussed in Ref.~\cite{Dvo14}.
The quantum states of a Dirac neutrino in matter and electromagnetic
fields were described in Ref.~\cite{ArbLobMur10}. The wave equation
for a neutrino in matter and a constant magnetic field was solved
in Ref.~\cite{StuTok14}. The exact solution for a neutrino with
anomalous magnetic moment interacting with matter and an electromagnetic
wave was found in Refs.~\cite{Dvo18,Dvo19}.

In the present work, we study the quantization of neutrinos interacting
with background matter and with a magnetic field. We assume that a
massive neutrino is a Dirac particle. Despite numerous attempts to
reveal the nature of neutrinos (see, e.g., Ref.\ \cite{Agr25}),
the issue whether neutrinos are Dirac or Majorana particles is still
open. We quantized massive Majorana neutrinos in vacuum in Ref.~\cite{Dvo12a}
and external fields in Refs.~\cite{Dvo12b,Dvo26} on the basis of
the Hamiltonian approach. In that formalism, the classical Majorana
field had a $c$-number wavefunction. The canonical quantization of
massive Majorana neutrinos in external fields was carried out in Ref.~\cite{DvoGit13}.

Here, we quantize massive Dirac neutrinos in background matter in
Sec.~\ref{sec:QUANTMATT}. The quantization of neutrinos interacting
with a magnetic field is carried out in Sec.~\ref{sec:QUANTB}. Using the results of Sec.~\ref{sec:QUANTB}, in Sec.~\ref{sec:PROPB}, we derive the propagator of a massive Dirac neutrino in a magnetic field. Finally,
we conclude in Sec.~\ref{sec:CONCL}.

\section{Quantization of a Dirac neutrino in matter}\label{sec:QUANTMATT}

The Lagrangian for a single massive neutrino interacting with background
matter reads
\begin{equation}\label{eq:Lagr}
  \mathcal{L}=\bar{\psi}
  \left[
    \mathrm{i}\gamma^{\mu}\partial_{\mu}-m-\frac{1}{2}g^{\mu}\gamma_{\mu}(1-\gamma^{5})
  \right]
  \psi,
\end{equation}
where $\gamma^{\mu}=(\gamma^{0},\bm{\gamma})$ and $\gamma^{5}=\mathrm{i}\gamma^{0}\gamma^{1}\gamma^{2}\gamma^{3}$
are the Dirac matrices, $m$ is the neutrino mass, and $g^{\mu}$
is the effective potential of the matter interaction. In nonmoving
and unpolarized matter, only $g^{0}\equiv g$ is nonzero. The values
of $g$ in the explicit form for different types of neutrinos can
be found in Ref.~\cite{DvoStu02}. The generalization of Eq.~(\ref{eq:Lagr})
for different neutrino types is straightforward.

In the classical field theory, one defines the canonical momentum
conjugate to $\psi$,
\begin{equation}
  \pi=\frac{\partial_{r}\mathcal{L}}{\partial(\partial_{t}\psi)}=\mathrm{i}\psi^{*},
\end{equation}
where $\partial_{r}$ is the right derivative since we take that $\psi$
is the Grassmann valued variable. The Poisson bracket of $\psi$ and
$\pi$ reads,
\begin{equation}\label{eq:Poisson}
  \left\{
    \psi(\mathbf{x},t),\pi(\mathbf{y},t)
  \right\}
  =\delta(\mathbf{x}-\mathbf{y}).
\end{equation}
When one quantizes the fermion in question, we replace $\psi$ and
$\pi$ with the operators which obey the equal time anticommutator,
\begin{equation}\label{eq:anticomgen}
  \left\{
    \psi(\mathbf{x},t),\pi(\mathbf{y},t)
  \right\} _{+}
  =\mathrm{i}\delta(\mathbf{x}-\mathbf{y}),
\end{equation}
instead of the Poisson bracket in Eq.~(\ref{eq:Poisson}).

To proceed, we write down the wave equation for $\psi$, resulting
from Eq.~(\ref{eq:Lagr}), as 
\begin{equation}\label{eq:Direqmattdiag}
  \mathrm{i}\dot{\psi}=H\psi,
  \quad
  H=\bm{\alpha}\mathbf{p}+\beta m+\frac{1}{2}g(1-\gamma^{5}),
\end{equation}
where $\bm{\alpha}=\gamma^{0}\bm{\gamma}$ and $\beta=\gamma^{0}$
are the Dirac matrices, and $\mathbf{p}=-\mathrm{i}\nabla$ is the
momentum operator.

One can check that the helicity operator $(\bm{\Sigma}\mathbf{p})$
commutes with the Hamiltonian of Eq.~(\ref{eq:Direqmattdiag}) provided
that $g$ is spatially uniform. Here, $\bm{\Sigma}=\gamma^{5}\bm{\alpha}$ are the Dirac matrices.
Hence, $\psi$ is eigenfunction of
$(\bm{\Sigma}\mathbf{p})$, $(\bm{\Sigma}\mathbf{p})\psi=\sigma p\psi$,
where $\sigma=\pm$. The general solution of Eq.~(\ref{eq:Direqmattdiag})
has the form,
\begin{align}\label{eq:psiamatt}
  \psi(\mathbf{x},t)= & \int\frac{\mathrm{d}^{3}p}{(2\pi)^{3/2}}e^{-\mathrm{i}gt/2}
  \nonumber
  \\
  & \times
  \sum_{\sigma=\pm}
  \big(
    a_{\sigma}(\mathbf{p})u_{\sigma}(\mathbf{p})e^{-\mathrm{i}E_{\sigma}t+\mathrm{i}\mathbf{px}}
    \notag
    \\
    & +
    b_{\sigma}^{\dagger}(\mathbf{p})v_{\sigma}  (\mathbf{p})e^{\mathrm{i}E_{\sigma}t-\mathrm{i}\mathbf{px}}
  \big),
\end{align}
where
\begin{equation}\label{eq:enlevmatt}
  E_{\sigma}=\sqrt{\left(p-\sigma\frac{g}{2}\right)^{2}+m^{2}},
\end{equation}
are the energy levels,
\begin{align}\label{eq:basismatt}
  u_{\sigma}(\mathbf{p})= & \sqrt{\frac{E_{\sigma}-\sigma p+g/2}{2E_{\sigma}}}
  \nonumber
  \\
  & \times
  \left(
    \begin{array}{c}
      -\frac{m_{a}}{E_{\sigma}-\sigma p+g/2}w_{\sigma}(\mathbf{p})
      \\
      w_{\sigma}(\mathbf{p})
    \end{array}
  \right),
  \nonumber
  \\
  v_{\sigma}(\mathbf{p})= & \sqrt{\frac{E_{\sigma}-\sigma p+g/2}{2E_{\sigma}}}
  \nonumber
  \\
  & \times
  \left(
    \begin{array}{c}
      w_{-\sigma}(\mathbf{p})
      \\
      \frac{m}{E_{\sigma}-\sigma p+g/2}w_{-\sigma}(\mathbf{p})
    \end{array}
  \right),
\end{align}
are the basis spinors, $b_{\sigma}^{\dagger}(\mathbf{p})$ and $a_{\sigma}(\mathbf{p})$
are the creation and annihilation operators for antineutrinos and
neutrinos, and
\begin{align}\label{eq:helamp}
  w_{+}(\mathbf{p})= &
  \left(
    \begin{array}{c}
      e^{-\mathrm{i}\varphi/2}\cos\vartheta/2
      \\
      e^{\mathrm{i}\varphi/2}\sin\vartheta/2
    \end{array}
  \right),
  \nonumber
  \\
  w_{-}(\mathbf{p})= &
  \left(
    \begin{array}{c}
      -e^{-\mathrm{i}\varphi/2}\sin\vartheta/2
      \\
      e^{\mathrm{i}\varphi/2}\cos\vartheta/2
    \end{array}
  \right),
\end{align}
are the helicity amplitudes. Here, $\vartheta$ and $\varphi$ are
spherical angles fixing the direction of $\mathbf{p}$. Equation~(\ref{eq:basismatt})
implies that the Dirac matrices are in the chiral representation.

One can check that $w_{\sigma}$ satisfy the following relations:
\begin{equation}\label{eq:helampprop}
  w_{\sigma}^{\dagger}(\mathbf{p})w_{\sigma'}(\mathbf{p})=\delta_{\sigma\sigma'},
  \quad
  w_{\sigma}(-\mathbf{p})=\mathrm{i}w_{-\sigma}(\mathbf{p}),
\end{equation}
based on Eq.~(\ref{eq:helamp}). The basis spinors in Eq.~(\ref{eq:basismatt})
obey the conditions,
\begin{align}\label{eq:normcond}
  u_{\sigma}^{\dagger}(\mathbf{p})u_{\sigma'}(\mathbf{p}) &
  =v_{\sigma}^{\dagger}(\mathbf{p})v_{\sigma'}(\mathbf{p})=\delta_{\sigma\sigma'},
  \nonumber
  \\
  u_{\sigma}^{\dagger}(\mathbf{p})v_{\sigma'}(-\mathbf{p}) &
  =v_{\sigma}^{\dagger}(\mathbf{p})u_{\sigma'}(-\mathbf{p})=0,
\end{align}
which can be verified using Eq.~(\ref{eq:helampprop}).

In Ref.~\cite{Dvo25c}, we checked that the equal time anticommutator
$\left\{ \psi(\mathbf{x},t),\psi^{\dagger}(\mathbf{y},t)\right\} _{+}$
of $\psi$ given in Eq.~(\ref{eq:psiamatt}) is consistent with Eq.~(\ref{eq:anticomgen})
provided that the creation and annihilation operators satisfy the
canonical anticommutation relations,
\begin{align}\label{eq:anticom}
  \left\{
    a_{\sigma}(\mathbf{p}),a_{\sigma'}^{\dagger}(\mathbf{q})
  \right\} _{+} & =
  \left\{
    b_{\sigma}(\mathbf{p}),b_{\sigma'}^{\dagger}(\mathbf{q})
  \right\} _{+}
  \nonumber
  \\
  & =\delta_{\sigma\sigma'}\delta(\mathbf{p}-\mathbf{q}),
\end{align}
with the rest of the anticommutators being equal to zero. Now, we
have to check that the total energy and the momentum of $\psi$ have
the appropriate forms.

The total energy of a massive neutrino is $E=\smallint\mathrm{d}^{3}x\psi^{\dagger}H\psi$.
Using Eq.~(\ref{eq:Direqmattdiag}), we rewrite $E$ in the form,
\begin{equation}\label{eq:Edef}
  E=\mathrm{i}\int\mathrm{d}^{3}x\psi^{\dagger}\frac{\partial\psi}{\partial t}.
\end{equation}
Based on Eqs.~(\ref{eq:psiamatt}), (\ref{eq:basismatt}), and~(\ref{eq:normcond}),
we get the following expression for $E$ in Eq.~(\ref{eq:Edef}):
\begin{align}\label{eq:Eexc}
  E= & \int\mathrm{d}^{3}p\sum_{\sigma=\pm}
  \big[
    (E_{\sigma}+g/2)a_{\sigma}^{\dagger}(\mathbf{p})a_{\sigma}(\mathbf{p})
    \notag
    \\
    & +
    (E_{\sigma}-g/2)b_{\sigma}^{\dagger}(\mathbf{p})b_{\sigma}(\mathbf{p})
  \big]
  \notag
  \\
  & +
  \text{vacuum terms}.
\end{align}
The total momentum of a massive neutrino is $\mathbf{P}=-\mathrm{i}\smallint\mathrm{d}^{3}x\psi^{\dagger}\nabla\psi$.
Analogously to the total energy, the expression for $\mathbf{P}$
takes the form,
\begin{align}\label{eq:Pexc}
  \mathbf{P}= & \int\mathrm{d}^{3}p\mathbf{p}\sum_{\sigma=\pm}
  \left[
    a_{\sigma}^{\dagger}(\mathbf{p})a_{\sigma}(\mathbf{p})+b_{\sigma}^{\dagger}(\mathbf{p})b_{\sigma}(\mathbf{p})
  \right]
  \nonumber
  \\
  & +
  \text{vacuum terms}.
\end{align}
One can see in Eqs.~(\ref{eq:Eexc}) and~(\ref{eq:Pexc}) that the
total energy and momentum of $\psi$ is the sum of contributions of
independent oscillators if we neglect divergent vacuum terms. Based
on Eqs.~(\ref{eq:Eexc}) and~(\ref{eq:Pexc}), in Table~\ref{tab:quantnum},
we list the quantum numbers of the elementary excitations of the massive
neutrino field in background matter. 

\begin{table*}
\renewcommand{\arraystretch}{1.25}
\renewcommand{\tabcolsep}{3pt}
\begin{center}\caption{The helicity, the momentum, and the energy of the elementary excitations
of the massive neutrino field in background matter.\label{tab:quantnum}}
\begin{tabular}{|c|c|c|c|}
\hline 
Particle type & Helicity, $\sigma$ & Momentum & Energy\tabularnewline
\hline 
Active neutrino & -1 & $\mathbf{p}$ & $E_{-}+g/2\approx p+g+\tfrac{m^{2}}{2p}$\tabularnewline
\hline 
Sterile neutrino & +1 & $\mathbf{p}$ & $E_{+}+g/2\approx p+\tfrac{m^{2}}{2p}$\tabularnewline
\hline 
Active antineutrino & +1 & $\mathbf{p}$ & $E_{+}-g/2\approx p-g+\tfrac{m^{2}}{2p}$\tabularnewline
\hline 
Sterile antineutrino & -1 & $\mathbf{p}$ & $E_{-}-g/2\approx p+\tfrac{m^{2}}{2p}$\tabularnewline
\hline 
\end{tabular}
\end{center}
\end{table*}

\section{Quantization of a Dirac neutrino in a magnetic field}\label{sec:QUANTB}

If a massive Dirac neutrino interacts with an electromagnetic field,
the Lagrangian has the form
\begin{equation}\label{eq:LagrB}
  \mathcal{L}=\bar{\psi}
  \left[
    \mathrm{i}\gamma^{\mu}\partial_{\mu}-m-\frac{\mu}{2}F_{\mu\nu}\sigma^{\mu\nu}
  \right]
  \psi,
\end{equation}
where $\sigma^{\mu\nu}=\tfrac{\mathrm{i}}{2}[\gamma^{\mu},\gamma^{\nu}]_{-}$
are the Dirac matrices, $\mu$ is the diagonal neutrino magnetic moment,
and $F_{\mu\nu}=(\mathbf{E},\mathbf{B})$ is the electromagnetic field
tensor. We write down the wave equation, resulting from Eq.~(\ref{eq:LagrB}),
in the Hamilton form,
\begin{equation}\label{eq:DireqBdiag}
  \mathrm{i}\dot{\psi}=H\psi,
  \quad
  H=\bm{\alpha}\mathbf{p}+\beta m-\mu B\beta\Sigma_{3},
\end{equation}
where we assume that the electric field is
absent, $\mathbf{E}=0$, and the magnetic field is along the $z$-axis,
$\mathbf{B}=B\mathbf{e}_{z}$.

The general solution of Eq.~(\ref{eq:DireqBdiag}) can be present
in the form,
\begin{align}\label{eq:psiaB}
  \psi(\mathbf{x},t) = & \int\frac{\mathrm{d}^{3}p}{(2\pi)^{3/2}}\sum_{\zeta=\pm}
  \Big(
    a_{\zeta}(\mathbf{p})u_{\zeta}(\mathbf{p})e^{-\mathrm{i}E_{\zeta}t}
    \notag
    \\
    & +
    b_{\zeta}^{\dagger}(\mathbf{p})v_{\zeta}(\mathbf{p})e^{\mathrm{i}E_{\zeta}t}
    \Big)
    e^{\mathrm{i}\mathbf{px}},
\end{align}
where $b_{\zeta}^{\dagger}(\mathbf{p})$ and $a_{\zeta}(\mathbf{p})$
are creation and annihilation operators. The energy levels in Eq~(\ref{eq:psiaB})
were obtained in Ref.\ \cite{TerBagKha65}, where a neutron in a
magnetic field was considered. They are
\begin{equation}\label{eq:EnB}
  E_{\zeta}=\sqrt{(S-\zeta\mu B)^{2}+p_{3}^{2}},
\end{equation}
where $S=\sqrt{m^{2}+p_{\perp}^{2}}$. The basis spinor corresponding
to particle degrees of freedom was also found in Ref.\ \cite{TerBagKha65},
\begin{equation}\label{eq:uzeta}
  u_{\zeta}=\frac{1}{2\sqrt{2}}
  \left(
    \begin{array}{c}
      z_{+}(f_{+}+\zeta f_{-})
      \\
      -\zeta z_{-}(f_{+}-\zeta f_{-})e^{\mathrm{i}\phi}
      \\
      z_{+}(f_{+}-\zeta f_{-})
      \\
      \zeta z_{-}(f_{+}+\zeta f_{-})e^{\mathrm{i}\phi}
    \end{array}
  \right),
\end{equation}
where
\begin{align}\label{eq:zf}
  z_{\pm} & =\sqrt{1\pm\zeta\frac{m}{S}},
  \quad
  \tan\phi =p_{1}/p_{2},
  \notag
  \\
  f_{\pm} & =\sqrt{1\pm\frac{p_{3}}{E_{\zeta}}}.
\end{align}
Here we take the Dirac matrices in the Dirac representation. In Eqs.~(\ref{eq:EnB})-(\ref{eq:zf}),
we assume the decomposition of the momentum $\mathbf{p}=\mathbf{p}_{\perp}+p_{3}\mathbf{e}_{z}$
and $\mathbf{p}_{\perp}=p_{1}\mathbf{e}_{x}+p_{2}\mathbf{e}_{y}$.

To complete the solution in Eq.~(\ref{eq:psiaB}), we provide the
detailed derivation of $v_{\zeta}(\mathbf{p})$. First, we notice
that the polarization operator
\begin{equation}
  \Pi=m\gamma^{0}-\mathrm{i}\gamma^{0}\gamma^{5}(\bm{\Sigma}\times\mathbf{p})_{3}-\mu B.
\end{equation}
commutes with the Hamiltonian in Eq.~(\ref{eq:DireqBdiag}). Thus,
$\Pi\psi=\zeta\lambda\psi$, where $\lambda$ is the eigenvalue of
$\Pi$ and $\zeta=\pm1$. Then, one finds that $\Pi^{2}=H^{2}-p_{3}^{2}$.
Hence $\lambda^{2}=E_{\zeta}^{2}-p_{3}^{2}$. To find $\lambda$,
we represent $v^{\mathrm{T}}=(C_{1},C_{2}e^{\mathrm{i}\phi},C_{3},C_{4}e^{\mathrm{i}\phi})$.
The eigenvalue equation for $\Pi$ becomes,
\begin{align}\label{eq:CiPi}
  (m-\mu B-\zeta\lambda)C_{1}+p_{\perp}C_{4} & =0,
  \nonumber 
  \\
  (m+\mu B+\zeta\lambda)C_{2}+p_{\perp}C_{3} & =0,
  \nonumber 
  \\
  (m-\mu B-\zeta\lambda)C_{3}-p_{\perp}C_{2} & =0,
  \nonumber 
  \\
  (m+\mu B+\zeta\lambda)C_{4}-p_{\perp}C_{1} & =0.
\end{align}
Equating the determinant of the system in Eq.~(\ref{eq:CiPi}) to
zero, one gets that $\lambda=S-\zeta\mu B$. Therefore, we reproduce
the energy spectrum in Eq.~(\ref{eq:EnB}).

Dirac Eq.~(\ref{eq:DireqBdiag}) for the negative energy spinors
becomes, $-Ev=Hv$, which can be rewritten in terms of the coefficients
$C_{i}$ as
\begin{align}\label{eq:CiDireq}
  (E+m-\mu B)C_{1} & =-p_{3}C_{3}+p_{\perp}C_{4},
  \nonumber
  \\
  (E+m+\mu B)C_{2} & =p_{3}C_{4}-p_{\perp}C_{3},
\end{align}
where we write down only two equations our of four. Using Eq.~(\ref{eq:CiPi}),
we cast Eq.~(\ref{eq:CiDireq}) to the form,
\begin{align}\label{eq:CiHPi}
  (E+\zeta\lambda)C_{1} & =-p_{3}C_{3},
  \nonumber
  \\
  (E-\zeta\lambda)C_{2} & =p_{3}C_{4}.
\end{align}
One can notice that the solution of Eq.~(\ref{eq:CiHPi}) is represented
as
\begin{align}
  C_{1} & =-\zeta F(f_{+}-\zeta f_{-}),
  \quad 
  C_{2}=G(f_{+}+\zeta f_{-}),
  \nonumber 
  \\
  C_{3} & =\zeta F(f_{+}+\zeta f_{-}),
  \quad 
  C_{4}=G(f_{+}-\zeta f_{-}),
\end{align}
where $F$ and $G$ are undetermined coefficients, and $f_{\pm}$
are given in Eq.~(\ref{eq:zf}).

To find $F$ and $G$, we use the normalization condition, $|v|^{2}=1$,
and the fact that $C_{3}/C_{2}=p_{\perp}/(m-\zeta S)$, which results
from Eq.~(\ref{eq:CiPi}). Eventually, one gets that $F=\zeta z_{+}/2\sqrt{2}$
and $G=-\zeta z_{-}/2\sqrt{2}$. Rescaling $-\zeta v\to v$, we obtain
that
\begin{equation}\label{eq:vzeta}
  v_{\zeta}=\frac{1}{2\sqrt{2}}
  \left(
    \begin{array}{c}
      \zeta z_{+}(f_{+}-\zeta f_{-})
      \\
      z_{-}(f_{+}+\zeta f_{-})e^{\mathrm{i}\phi}
      \\
      -\zeta z_{+}(f_{+}+\zeta f_{-})
      \\
      z_{-}(f_{+}-\zeta f_{-})e^{\mathrm{i}\phi}
    \end{array}
  \right),
\end{equation}
where $z_{\pm}$ are given in Eq.~(\ref{eq:zf}).

We mention the following properties of the basis spinors:
\begin{equation}\label{eq:normB}
  u_{\zeta}^{\dagger}u_{\zeta'}=v_{\zeta}^{\dagger}v_{\zeta'}=\delta_{\zeta\zeta'},
  \quad 
  u_{\zeta}^{\dagger}v_{\zeta'}=0,
\end{equation}
and
\begin{widetext}
\begin{align}\label{eq:sumzetaB}
  & \sum_{\zeta=\pm} (u_{\zeta}\otimes u_{\zeta}^{\dagger}) =
  \notag
  \\
  \notag
  &
  \frac{1}{8}
  \left(
    \begin{array}{cccc}
      Y_{-}^{2}+X_{+}^{2}-\frac{m}{S}(Y_{-}^{2}-X_{+}^{2}); &
      2e^{-\mathrm{i}\phi}\varDelta p_{\perp}/S; & 
      2\varSigma_{+}; & 
      e^{-\mathrm{i}\phi}\left(X_{-}^{2}-Y_{+}^{2}\right)p_{\perp}/S
      \\
      2z_{-}z_{+}e^{\mathrm{i}\phi}\varDelta; & 
      Y_{+}^{2}+X_{-}^{2}-\frac{m}{S}(Y_{+}^{2}-X_{-}^{2}); & 
      e^{\mathrm{i}\phi}\left(X_{+}^{2}-Y_{-}^{2}\right)p_{\perp}/S; & 
      -2\varSigma_{-}
      \\
      2\varSigma_{+}; & 
      e^{-\mathrm{i}\phi}\left(X_{+}^{2}-Y_{-}^{2}\right)p_{\perp}/S; & 
      X_{-}^{2}+Y_{+}^{2}-\frac{m}{S}(X_{-}^{2}-Y_{+}^{2}); & 
      -2e^{-\mathrm{i}\phi}\varDelta p_{\perp}/S
      \\
      e^{\mathrm{i}\phi}\left(X_{-}^{2}-Y_{+}^{2}\right)p_{\perp}/S; & 
      -2\varSigma_{-}; & 
      -2e^{\mathrm{i}\phi}\varDelta p_{\perp}/S; & 
      X_{+}^{2}+Y_{-}^{2}-\frac{m}{S}(X_{+}^{2}-Y_{-}^{2})
    \end{array}
  \right),
  \\
  \notag
  & \sum_{\zeta=\pm} (v_{\zeta}\otimes v_{\zeta}^{\dagger}) =
  \\
  &
  \frac{1}{8}
  \left(
    \begin{array}{cccc}
      X_{-}^{2}+Y_{+}^{2}-\frac{m}{S}(X_{-}^{2}-Y_{+}^{2}); & 
      -2e^{-\mathrm{i}\phi}\varDelta p_{\perp}/S; & 
      -2\varSigma_{+}; & 
      e^{-\mathrm{i}\phi}\left(Y_{+}^{2}-X_{-}^{2}\right)p_{\perp}/S
      \\
      -2e^{\mathrm{i}\phi}\varDelta p_{\perp}/S; & 
      X_{+}^{2}+Y_{-}^{2}-\frac{m}{S}(X_{+}^{2}-Y_{-}^{2}); & 
      e^{\mathrm{i}\phi}\left(Y_{-}^{2}-X_{+}^{2}\right)p_{\perp}/S; & 
      2\varSigma_{-}
      \\
      -2\varSigma_{+}; & 
      e^{-\mathrm{i}\phi}\left(Y_{-}^{2}-X_{+}^{2}\right)p_{\perp}/S; & 
      Y_{-}^{2}+X_{+}^{2}-\frac{m}{S}(Y_{-}^{2}-X_{+}^{2}); & 
      2e^{-\mathrm{i}\phi}\varDelta p_{\perp}/S
      \\
      e^{\mathrm{i}\phi}\left(Y_{+}^{2}-X_{-}^{2}\right)p_{\perp}/S; & 
      2\varSigma_{-}; & 
      2e^{\mathrm{i}\phi}\varDelta p_{\perp}/S; & 
      Y_{+}^{2}+X_{-}^{2}-\frac{m}{S}(Y_{+}^{2}-X_{-}^{2})
    \end{array}
  \right),
\end{align}
\end{widetext}
which can be checked by means of the direct calculations based on
Eqs.~(\ref{eq:uzeta}) and~(\ref{eq:vzeta}). In Eq.~(\ref{eq:sumzetaB}),
we use the following notations:
\begin{align}
  X_{\pm} & =\sqrt{1-\frac{p_{3}}{E_{\pm}}}+\sqrt{1+\frac{p_{3}}{E_{\pm}}},
  \nonumber 
  \\
  Y_{\pm} & =\sqrt{1-\frac{p_{3}}{E_{\pm}}}-\sqrt{1+\frac{p_{3}}{E_{\pm}}},
  \nonumber 
  \\
  \varDelta & =p_{3}\left(\frac{1}{E_{-}}-\frac{1}{E_{+}}\right),
  \nonumber 
  \\
  \varSigma_{\pm} & =p_{3}\left(\frac{z_{-}^{2}}{E_{\mp}}+\frac{z_{+}^{2}}{E_{\pm}}\right),
\end{align}
which are the dimensionless quantities.

To quantize Dirac neutrinos in a magnetic field we should, first,
check that $\left\{ \psi(\mathbf{x},t),\psi^{\dagger}(\mathbf{y},t)\right\} _{+}=\delta(\mathbf{x}-\mathbf{y})$
for the decomposition in Eq.~(\ref{eq:psiaB}), provided that the
creation and annihilation operators obey the canonical anticommutation
relations in Eq.~(\ref{eq:anticom}). One gets from Eq.~(\ref{eq:psiaB})
that
\begin{multline}
  \left\{
    \psi(\mathbf{x},t),\psi^{\dagger}(\mathbf{y},t)
  \right\} _{+}= 
  \int\frac{\mathrm{d}^{3}p}{(2\pi)^{3}}e^{\mathrm{i}\mathbf{p}(\mathbf{x}-\mathbf{y})}
  \\
  \times
  \sum_{\zeta=\pm}
  \left[
    (u_{\zeta}\otimes u_{\zeta}^{\dagger})+(v_{\zeta}\otimes v_{\zeta}^{\dagger})
  \right]
  =\delta(\mathbf{x}-\mathbf{y}),
\end{multline}
where we take into account the fact that
\begin{equation}
  \sum_{\zeta=\pm}
  \left[
    (u_{\zeta}\otimes u_{\zeta}^{\dagger})+(v_{\zeta}\otimes v_{\zeta}^{\dagger})
  \right]
  =1,
\end{equation}
which results from Eq.~(\ref{eq:sumzetaB}).

The total energy and the total momentum of the Dirac neutrino field
are computed in the same manner as in Sec.~\ref{sec:QUANTMATT}.
We just provide the final expressions for the energy
\begin{align}\label{eq:EexcB}
  E = & \int\mathrm{d}^{3}p
  \sum_{\zeta=\pm}E_{\zeta}
  \left[
    a_{\zeta}^{\dagger}(\mathbf{p})a_{\zeta}(\mathbf{p})+b_{\zeta}^{\dagger}(\mathbf{p})b_{\zeta}(\mathbf{p})
  \right]
  \notag
  \\
  & +
  \text{vacuum terms},
\end{align}
and for the total momentum,
\begin{align}\label{eq:PexcB}
  \mathbf{P}= & \int\mathrm{d}^{3}p\mathbf{p}
  \sum_{\zeta=\pm}
  \left[
    a_{\zeta}^{\dagger}(\mathbf{p})a_{\zeta}(\mathbf{p})-b_{\zeta}^{\dagger}(\mathbf{p})b_{\zeta}(\mathbf{p})
  \right]
  \notag
  \\
  & +\text{vacuum terms},
\end{align}
where we use Eq.~(\ref{eq:normB}) and the proper anticommutation
relations for operators.

The expression for the energy in Eq.~(\ref{eq:EexcB}) has the appropriate
form. The contribution of the negative energy degrees of freedom to
the total momentum in Eq.~(\ref{eq:PexcB}) has the negative sign.
It results from the noncovariant phase of the wave function part in
Eq.~(\ref{eq:psiaB}) corresponding to these degrees of freedom:
$E_{\zeta}t+\mathbf{px}$ vs. $E_{\zeta}t-\mathbf{px}$. Thus, the
operators $b_{\zeta}(\mathbf{p})$ and $b_{\zeta}^{\dagger}(\mathbf{p})$
are attributed to holes rather than to antiparticles. The momentum
of an antineutrino is opposite to that of a hole: $\mathbf{p}_{\bar{\nu}}=-\mathbf{p}_{\mathrm{hole}}$.

To illustrate the values of the quantum numbers of a massive Dirac
neutrino in a magnetic field, we consider the situation when a particle
propagates along the $x$-axis and assume that the magnetic field
is transverse: $\mathbf{B}=B\mathbf{e}_{z}$. In this situation, $\mathbf{p}_{\perp}=p\mathbf{e}_{x}$,
$p_{3}=0$, and $S=\sqrt{p^{2}+m^{2}}\equiv E_{0}$ is the energy
of a noninteracting particle. We list the values for the energy and
the momentum of neutrinos, holes, and antineutrinos in Table~\ref{tab:quantnumB}.

\begin{table*}
\renewcommand{\arraystretch}{1.25}
\renewcommand{\tabcolsep}{3pt}
\begin{center}
\caption{The momentum and the energy of the elementary excitations of a massive
Dirac neutrino propagating in a transverse magnetic field.\label{tab:quantnumB}}
\begin{tabular}{|c|c|c|}
\hline 
Particle type & Momentum & Energy\tabularnewline
\hline 
Neutrinos & $p\mathbf{e}_{x}$ & $E_{0}\mp\mu B\approx p+\tfrac{m^{2}}{2p}\mp\mu B$\tabularnewline
\hline 
Holes & $-p\mathbf{e}_{x}$ & $E_{0}\mp\mu B\approx p+\tfrac{m^{2}}{2p}\mp\mu B$\tabularnewline
\hline 
Antineutrinos & $p\mathbf{e}_{x}$ & $E_{0}\mp\mu B\approx p+\tfrac{m^{2}}{2p}\mp\mu B$\tabularnewline
\hline 
\end{tabular}
\end{center}
\end{table*}

\section{Propagator of a Dirac neutrino in a magnetic field}\label{sec:PROPB}

Now, we use the results of Sec.~\ref{sec:QUANTB} to derive the propagator
of a massive Dirac neutrino in a magnetic field. The propagator $S(x)$
of a fermion is defined in the common way,
\begin{align}\label{eq:propdef}
  \mathrm{i}S(x-y) = & \theta(x_{0}-y_{0})
  \left\langle 0\left|
    \psi(x)\bar{\psi}(y)
  \right|0\right\rangle 
  \notag
  \\
  & -
  \theta(y_{0}-x_{0})
  \left\langle 0\left|
    \bar{\psi}(y)\psi(x)
  \right|0\right\rangle,
\end{align}
where $\theta(t)$ is the Heaviside step function. In Eq.~(\ref{eq:propdef}),
the wavefunction $\psi(x)$ is given in Eq.~(\ref{eq:psiaB}) which
accounts for the contribution of the magnetic field.

The vacuum averagings in Eq.~(\ref{eq:psiaB}) are computed using the properties
of the creation and annihilation operators,
\begin{align}\label{eq:vacav}
  \left\langle 0\left|
    \psi(x)\bar{\psi}(y)
  \right|0\right\rangle = &
  \int\frac{\mathrm{d}^{3}p}{(2\pi)^{3}} e^{\mathrm{i}\mathbf{p}(\mathbf{x}-\mathbf{y})}
  \notag
  \\
  & \times
  \sum_{\zeta=\pm}e^{-\mathrm{i}E_{\zeta}(x_{0}-y_{0})}(u_{\zeta}\otimes\bar{u}_{\zeta}),
  \nonumber
  \\
  \left\langle 0\left|
    \bar{\psi}(y)\psi(x)
  \right|0\right\rangle  = &
  \int\frac{\mathrm{d}^{3}p}{(2\pi)^{3}} e^{\mathrm{i}\mathbf{p}(\mathbf{x}-\mathbf{y})}
  \notag
  \\
  & \times
  \sum_{\zeta=\pm}e^{\mathrm{i}E_{\zeta}(x_{0}-y_{0})}(v_{\zeta}\otimes\bar{v}_{\zeta}).
\end{align}
With help of the identity
\begin{equation}
  \frac{1}{2\pi\mathrm{i}}\int_{-\infty}^{+\infty}\frac{e^{-\mathrm{i}p_{0}t}\mathrm{d}p_{0}}{p_{0}\mp E_{\zeta}\pm\mathrm{i}0}
  =\mp e^{\mp\mathrm{i}E_{\zeta}t}\theta(\pm t),
\end{equation}
where $\mathrm{i}0$ is the small imaginary term, as well as accounting
for Eq.~(\ref{eq:vacav}), we cast the propagator in Eq.~(\ref{eq:propdef})
to the form,
\begin{align}\label{eq:Sxinter}
  S(x)= & -\mathrm{i}\int\frac{\mathrm{d}^{3}p}{(2\pi)^{3}}e^{\mathrm{i}\mathbf{p}\mathbf{x}}
  \sum_{\zeta=\pm}
  \big[
    (u_{\zeta}\otimes\bar{u}_{\zeta})e^{-\mathrm{i}E_{\zeta}t}\theta(t)
    \notag
    \\
    & -
    (v_{\zeta}\otimes\bar{v}_{\zeta})e^{\mathrm{i}E_{\zeta}t}\theta(-t)
  \big]
  \nonumber \\
 & =\int\frac{\mathrm{d}^{4}p}{(2\pi)^{4}}e^{-\mathrm{i}px}
 \notag
 \\
 & \times
 \sum_{\zeta=\pm}
 \left[
   \frac{(u_{\zeta}\otimes\bar{u}_{\zeta})}{p_{0}-E_{\zeta}+\mathrm{i}0}
   +\frac{(v_{\zeta}\otimes\bar{v}_{\zeta})}{p_{0}+E_{\zeta}-\mathrm{i}0}
  \right],
\end{align}
where the tensor products have to be computed using Eqs.~(\ref{eq:uzeta})
and~(\ref{eq:vzeta}).

Finally, based on Eq.~\eqref{eq:Sxinter}, one gets the 4D Fourier image of the propagator, $S(p)=\smallint\mathrm{d}^{4}xS(x)e^{\mathrm{i}px}$,
as
\begin{widetext}
\begin{equation}\label{eq:FimS}
  S(p)=\frac{1}{8}\sum_{\zeta=\pm}
  \left(
    \begin{array}{cccc}
      z_{+}^{2}A_{+}; & 
      -2\Delta\frac{\zeta z_{-}z_{+}p_{3}}{E_{\zeta}}e^{-\mathrm{i}\phi}; & 
      -2\Delta\frac{z_{+}^{2}p_{3}}{E_{\zeta}}; & -\zeta z_{-}z_{+}A_{+}e^{-\mathrm{i}\phi}
      \\
      -2\Delta\frac{\zeta z_{-}z_{+}p_{3}}{E_{\zeta}}e^{\mathrm{i}\phi}; & 
      z_{-}^{2}A_{-}; & \zeta z_{-}z_{+}A_{-}e^{\mathrm{i}\phi}; & 
      2\Delta\frac{z_{-}^{2}p_{3}}{E_{\zeta}}
      \\
      2\Delta\frac{z_{+}^{2}p_{3}}{E_{\zeta}}; & -\zeta z_{-}z_{+}A_{-}e^{-\mathrm{i}\phi}; & 
      -z_{+}^{2}A_{-}; & 
      -2\Delta\frac{\zeta z_{-}z_{+}p_{3}}{E_{\zeta}}e^{-\mathrm{i}\phi}
      \\
      \zeta z_{-}z_{+}A_{+}e^{\mathrm{i}\phi}; & 
      -2\Delta\frac{z_{-}^{2}p_{3}}{E_{\zeta}}; & 
      -2\Delta\frac{\zeta z_{-}z_{+}p_{3}}{E_{\zeta}}e^{\mathrm{i}\phi}; & 
      -z_{-}^{2}A_{+}
    \end{array}
  \right),
\end{equation}
\end{widetext}
where
\begin{align}
  A_{\pm} & =\frac{(f_{+}\pm\zeta f_{-})^{2}}{p_{0}-E_{\zeta}+\mathrm{i}0}+\frac{(f_{+}\mp\zeta f_{-})^{2}}{p_{0}+E_{\zeta}-\mathrm{i}0},
  \nonumber
  \\
  \Delta & =\frac{1}{p_{0}-E_{\zeta}+\mathrm{i}0}-\frac{1}{p_{0}+E_{\zeta}-\mathrm{i}0}.
\end{align}
Here, $z_{\pm}$ and $f_{\pm}$ are given in Eq.~(\ref{eq:zf}).
If we consider the limit $B\to0$, one can demonstrate that $S(p)$
in Eq.~(\ref{eq:FimS}) becomes
\begin{equation}
  S(p)\to\frac{\gamma^{\mu}p_{\mu}+m}{p^{2}-m^{2}+\mathrm{i}0}.
\end{equation}
That is, the propagator coincides with the
known expression for a fermion in vacuum.

\section{Conclusion}\label{sec:CONCL}

In this work, we have quantized a massive Dirac neutrino in various
external fields. We have considered two particular external backgrounds:
the electroweak interaction with matter and the interaction with a
magnetic field. In both situations, we have obtained the exact solutions
of classical wave equations which provided us with the energy levels
and the basis spinors for positive and negative energy states.

Then, we have checked that the equal time anticommutators of the operator
valued wave functions have the canonical forms. The total energy and
momentum of the neutrino field were shown to be present as the contributions
of independent oscillators. In case of the magnetic interaction considered
in Sec.~\ref{sec:QUANTB}, we have obtained the the negative energy
part of the neutrino wave function in Eq.~(\ref{eq:psiaB}) describes
holes rather than antiparticles.

The results obtained in the present work are of importance for the
quantum field theory based approach for neutrino oscillations. If
massive neutrinos interact with external fields while traveling from
a source to a detector, the propagators of these particles are based
on the exact solutions of wave equations accounting for external fields
(see, e.g., Refs.~\cite{Dvo25c,Dvo25a,Dvo25b}). The construction
of a neutrino propagator requires the properly quantized massive neutrino
field in an external background. For this purpose, in Sec.~\ref{sec:PROPB}, we have obtained the propagator of a massive Dirac neutrino exactly accounting for the contribution of a constant and uniform magnetic field.

\section*{CONFLICT OF INTEREST}
The author declares that there are no conflicts of interest.

\nocite{*}


\end{document}